\documentclass[epj,widetext]{svjour}
\usepackage{graphicx}  
\usepackage{dcolumn}   
\usepackage{bm}        
\usepackage{amssymb, amsmath}   
\usepackage[usenames]{color}
\usepackage[colorlinks,urlcolor=blue,citecolor=blue,linkcolor=blue]{hyperref}
\usepackage{cite}

\begin{document}
\newcommand{\bra}[1]{\mbox{\ensuremath{\langle #1 \vert}}}
\newcommand{\ket}[1]{\mbox{\ensuremath{\vert #1 \rangle}}}
\newcommand{\mb}[1]{\mathbf{#1}}
\newcommand{\phipp}{\big|\phi_{\mb{p}}^{(+)}\big>}
\newcommand{\phipav}{\big|\phi_{\mb{p}}^{\p{av}}\big>}
\newcommand{\pp}[1]{\big|\psi_{p}(#1)\big>}
\newcommand{\drdy}[1]{\sqrt{-R'(#1)}}
\newcommand{\Rb}{$^{87}$Rb}
\newcommand{\kf}{$^{40}$K}
\newcommand{\na}{${^{23}}$Na}
\newcommand{\muK}{\:\mu\textrm{K}}
\newcommand{\p}[1]{\textrm{#1}}
\newcommand\T{\rule{0pt}{2.6ex}}
\newcommand\B{\rule[-1.2ex]{0pt}{0pt}}
\newcommand{\reffig}[1]{\mbox{Fig.~\ref{#1}}}
\newcommand{\refeq}[1]{\mbox{Eq.~(\ref{#1})}}
\hyphenation{Fesh-bach}
\newcommand{\previous}[1]{}
\newcommand{\reddia}{\textcolor{red}{$\Diamond$}}
\newcommand{\gbox}{\textcolor{ForestGreen}{$\Box$}}
\newcommand{\bcirc}{\textcolor{blue}{$\bigcirc$}}

\setlength{\tabcolsep}{3pt}

\title{Feshbach resonances in the $^6$Li-$^{40}$K Fermi-Fermi mixture:\\ Elastic versus inelastic interactions}
\author{D. Naik\inst{1}
\and A. Trenkwalder\inst{1}
\and C. Kohstall\inst{1,2}
\and F. M. Spiegelhalder\inst{1}
\and M. Zaccanti\inst{1,3}
\and G. Hendl\inst{1}
\and F. Schreck\inst{1}
\and R.~Grimm\inst{1,2}                     
\and T. M. Hanna\inst{4}
\and P. S. Julienne\inst{4}} %
\institute{Institut f\"ur Quantenoptik und Quanteninformation,
\"Osterreichische Akademie der Wissenschaften, 6020 Innsbruck,
Austria
\and
Institut f\"ur Experimentalphysik, Universit\"at Innsbruck,
6020 Innsbruck, Austria
\and
LENS, Physics Department, University of Florence and INO-CNR, 50019 Sesto Fiorentino, Italy
\and
Joint Quantum Institute,
	NIST and University of Maryland,
	100 Bureau Drive, Stop 8423,
  	Gaithersburg MD 20899-8423, USA}
\date{Received: 18 October 2010}
%
\abstract{We present a detailed theoretical and experimental study of Feshbach resonances in the $^6$Li-$^{40}$K mixture.
Particular attention is given to the inelastic scattering properties, which have not been considered before. As an important example, we thoroughly investigate both elastic and inelastic scattering properties of a resonance that occurs near 155\,G. Our theoretical predictions based on a coupled channels calculation are found in excellent agreement with the experimental results.
We also present theoretical results on the molecular state that underlies the 155\,G resonance, in particular concerning its lifetime against spontaneous dissociation. We then present a survey of resonances in the system, fully characterizing the corresponding elastic and inelastic scattering properties. This provides the essential information to identify optimum resonances for applications relying on interaction control in this Fermi-Fermi mixture.} 
\maketitle

\section{Introduction}
A new frontier in the research field of strongly interacting Fermi gases \cite{Inguscio2006ufg,Giorgini2008tou} has been approached by the recent realizations of ultracold Fermi-Fermi mixtures of $^6$Li and $^{40}$K \cite{Taglieber2008qdt,Wille2008eau,Voigt2009uhf,Spiegelhalder2009cso,Tiecke2010bfr,Spiegelhalder2010aopNOTE,Costa2010swi}. Degenerate Fermi-Fermi mixtures represent a starting point to experimentally explore a rich variety of intriguing phenomena, such as many-body quantum phases of fermionic mixtures \cite{Liu2003igs,Forbes2005scf,Paananen2006pia,Iskin2006tsf,Iskin2007sai,Petrov2007cpo,Iskin2008tif,Baranov2008spb,Bausmerth2009ccl,Nishida2009ipw,Wang2009qpd,Mora2009gso,Gezerlis2009hlf,Diener2010bbc,Baarsma2010pam} and few-body quantum states \cite{Petrov2005dmi,Nishida2009cie,Levinsen2009ads,Nishida2010poa}.

The possibility to precisely tune the interspecies interaction via Feshbach resonances \cite{Chin2010fri} is an important prerequisite for many experiments. This has motivated theoretical and experimental work on Feshbach resonances in the $^6$Li-$^{40}$K mixture \cite{Wille2008eau,Tiemann2009cso,Tiecke2010bfr}. It turned out that all resonances for $s$-wave scattering in this system are quite narrow, the broadest ones exhibiting a width of $\lesssim$\,2\,G\footnote{Here, G\,$=0.1$\,mT.}, and their character is closed-channel dominated \cite{Chin2010fri}. This causes both practical and fundamental limitations for experimental applications. Interaction control is practically limited by magnetic field uncertainties and, on more fundamental grounds, the universal range \cite{Chin2010fri} near the center of the resonance is quite narrow.

Our work is motivated by identifying the Feshbach resonances in the $^6$Li-$^{40}$K system that are best suited for realizing Fermi-Fermi mixtures in the strongly interacting regime. In a previous study \cite{Tiecke2010bfr}, Tiecke {\em et al.} approached this question by calculating the widths of the different resonances \cite{Tiemann}, and they studied elastic scattering for one of the widest resonances in the system. Another important criterion is stability against inelastic two-body decay. For the $^6$Li-$^{40}$K mixture, inelastic spin-exchange collisions do not occur when at least one of the species is in its lowest spin state \cite{Wille2008eau}. When one of the species is in a higher state, decay is energetically possible, but rather weak as it requires spin-dipole coupling to outgoing higher partial waves. The wider resonances in the $^6$Li-$^{40}$K system are found in higher spin channels. This raises the important issue of possible inelastic two-body losses. The question of which is the optimum resonance for a particular application can only be answered if both width and decay are considered.

In this Article, we present a detailed study of Feshbach resonances in the $^6$Li-$^{40}$K system, characterizing their influence on both elastic and inelastic scattering properties. In Sec.\ \ref{sec:FRdecay} we briefly review a general formalism to describe decaying resonances. In Sec.\ \ref{sec:155G} we present a case study of a particularly interesting resonance. In theory and experiment, we investigate its elastic and inelastic scattering properties and the properties of the underlying molecular state. In Sec.\ \ref{sec:survey} we present a survey of all resonances, summarizing their essential properties. In Sec.\ \ref{sec:conclusion} we conclude by discussing the consequences of our insights for ongoing experiments towards strongly interacting Fermi-Fermi mixtures.

\section{Feshbach resonances with decay}
\label{sec:FRdecay}

A Feshbach resonance results from the coupling of a colliding atom pair to a near-degenerate bound state. If this molecular state can decay into open channels other than that in which the colliding pair is initially prepared, the situation is referred to as a decaying resonance \cite{Chin2010fri}. A formalism for describing such resonances has been developed for optical coupling~\cite{Fedichev1996ion,Bohn1999sto}, and has also been applied to the magnetically tunable case~\cite{Hutson2007fri}.
A well known example of a decaying resonance exists in the collision of two $^{85}$Rb atoms~\cite{Roberts1998rmf}, for which molecular lifetimes have been studied~\cite{Thompson2005sdo,Kohler2005sdo}.

The scattering properties in the zero-energy limit can be expressed through a complex $s$-wave scattering length, $\tilde{a} = a - ib$, where $a$ and $b$ are real. The relation of these two parameters to the two experimentally relevant quantities, the elastic scattering cross section $\sigma$ and the loss rate coefficient $K_2$ for inelastic decay, is for non-identical particles given by
\begin{equation}
\sigma = 4 \pi (a^2 + b^2)
\end{equation}
and
\begin{equation}
K_2 = \frac{2 h}{\mu} b \, .
\label{eq:K2b}
\end{equation}
Here, $h$ is Planck's constant and $\mu$ is the reduced mass.

The complex scattering length can be parametrized by
\begin{eqnarray}
a(B) &= &a_\p{bg}  -
a_{\rm{res}}\frac{\gamma_{\p{B}}(B - B_0)}
	{(B - B_0)^2 + (\gamma_\p{B}/2)^2}  \, ,
	\label{eq:aofb}
	\\
b(B) &= &2 a_\p{res}
\frac{(\gamma_\p{B}/2)^2}
{(B - B_0)^2 + (\gamma_{\rm{B}}/2)^2} \, .
\label{eq:bofb}
\end{eqnarray}

Here, $B$ is the magnetic field strength, the resonance occurs at $B = B_0$, and $a_\p{bg}$ is the background scattering length.
The decay of the ``bare'' molecular state that causes the resonance is characterized by a rate $\gamma$ \cite{Kohler2005sdo}, which we conveniently express in magnetic field units, $\gamma_\p{B} = \hbar \gamma/\delta\mu$, where $\delta\mu$ is the difference in magnetic moment between the entrance channel and the bare molecular state. The resonance length parameter $a_\p{res}$ is related to the resonance width $\Delta$ by
\begin{equation}
a_\p{res} = \frac{a_\p{bg} \Delta}{\gamma_\p{B}} \, ,
\end{equation}
and gives the range $a_\p{bg} \pm a_\p{res}$ within which the real part of the scattering length can vary, thus providing an indication of the possible control.
A common figure of merit for the coherent control of an ultracold gas is the ratio $a/b$.
For $|B - B_0| \gg \gamma_\p{B}$, and a change in scattering length much larger than $a_\p{bg}$, this can be shown from Eqs.~(\ref{eq:aofb}) and (\ref{eq:bofb}) to be
\begin{equation}
\frac{a}{b} \approx - 2 \frac{(B - B_0)}{\gamma_\p{B}}
\approx 2 \frac{a_\p{res}}{a}
\, .
\label{eq:asquared}
\end{equation}
A larger $a_\p{res}$ therefore gives better coherent control and lower losses for a given change in scattering length.
Combining Eqs.\ (\ref{eq:K2b}) and (\ref{eq:asquared}) gives a simple expression for the loss rate coefficient
\begin{equation}
K_2 \approx \frac{h}{\mu} \frac{a^2}{a_\p{res}} \, ,
\label{eq:K2a2}
\end{equation}
which shows a general $a^2$-scaling of two-body loss near a decaying Feshbach resonance.

\section{Case study of the 155\,G resonance}
\label{sec:155G}

In this Section, we present a thorough study of the 155\,G Feshbach resonance, which serves as our main tool for interaction tuning in strongly interacting Fermi-Fermi mixtures. It was first observed in Ref.\ \cite{Wille2008eau} and used for molecule formation in Ref.\ \cite{Voigt2009uhf}. We first (Sec.\ \ref{sec:sctheory}) present theoretical predictions for the elastic and inelastic scattering properties near this resonance based on coupled channels calculations. We then (Sec.\ \ref{sec:scexpts}) present our corresponding experimental results, providing a full confirmation of the expected resonance properties. We finally (Sec.\ \ref{sec:boundstate}) discuss the properties of the molecular state that causes the resonance.

\begin{figure}[tb]
	\centering
	\includegraphics[width=0.95\columnwidth, clip]{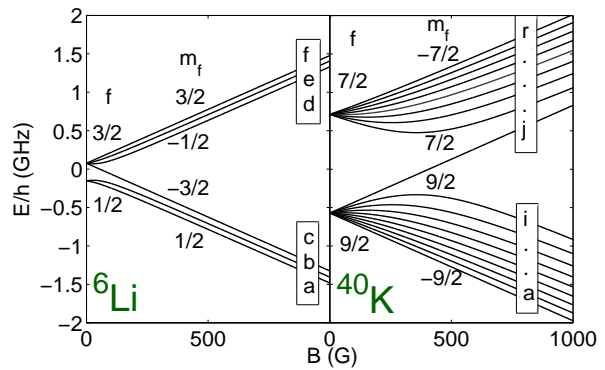}
	\caption{Zeeman sub-levels in the electronic ground state of $^{6}$Li and $^{40}$K, giving the total angular momentum $f$ and its projection $m_\p{f}$ along the quantization axis. We label Zeeman states alphabetically in order of increasing energy, as shown.}
	\label{fig:zeeman}
\end{figure}
Figure \ref{fig:zeeman} shows different magnetic and hyperfine sub-levels of the electronic ground states of $^6$Li and $^{40}$K. Here we follow the notation of Ref.\ \cite{Chin2010fri} and label the sub-states alphabetically in increasing order of energy. The 155\,G resonance occurs in the $ac$ channel, i.e.\ for a $^6$Li atom in state $a$ colliding with a $^{40}$K atom in state $c$.

\subsection{Scattering properties: Theory}
\label{sec:sctheory}

We have carried out coupled channels studies~\cite{Mies1996ebo} of the scattering properties in the $ac$ $s-$wave channel. The potentials used were taken from Ref.~\cite{Tiemann2009cso} and, to make this paper self-contained, we have summarized the important parameters in Table~\ref{table:params}.
\begin{table}[tb]
\begin{center}
\begin{tabular}{| c | c |}
\hline\T
singlet scattering length $a_\p{s}$ & $52.61\,a_0$ \\  \T
triplet scattering length $a_\p{t}$ & $64.41\,a_0$ \\ \hline \T
vdW coefficient $C_6$ & 2322\,$E_h$ \\ \T
vdW length  $R_\p{vdW} = \tfrac{1}{2} (2 \mu C_6/\hbar^2)^{1/4}$ &  40.8\,$a_0$ \\ \T
vdW energy $E_\p{vdW} = \hbar^2/(2\mu R_\p{vdW}^2)$ & $ h \times 207.6$\,MHz \\
\hline
\end{tabular}
\end{center}
\caption{Important parameters of the interaction potentials of $^{6}$Li--$^{40}$K, taken from the potentials of Ref.~\cite{Tiemann2009cso}. The van der Waals (vdW) parameters describe the long range part of the potential, $-C_6 / r^6$, where $r$ is the interparticle distance. Here, $a_0 = 0.529177 \times 10^{-10}$\,m is the Bohr radius and $E_h = 4.359744 \times 10^{-18}$\,J represents a hartree.}
\label{table:params}
\end{table}
\begin{figure}[tb]
	\centering
	\includegraphics[width=0.95\columnwidth, clip]{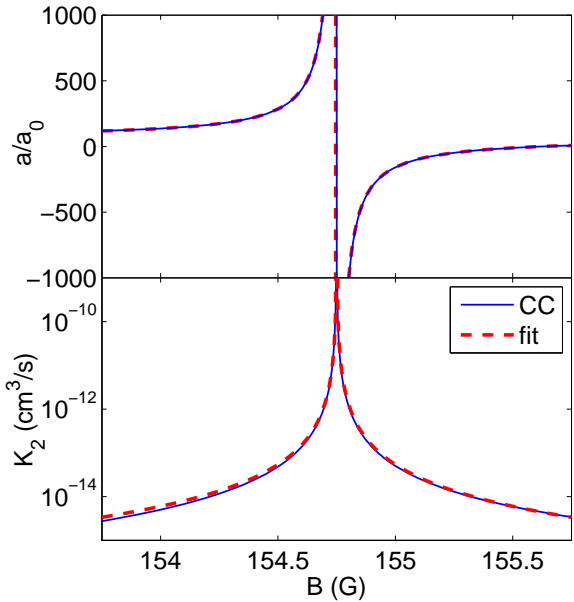}
	\caption{(Color online) $s$-wave scattering properties of the $ac$ channel, as a function of magnetic field.
	A coupled channels calculation (CC, solid line) is compared to a fit using Eqs.~(\ref{eq:K2b})--(\ref{eq:bofb}) (dashed line).
	The top panel shows the real part of the scattering length, while the two-body loss rate coefficient is shown in the lower panel.
	}
	\label{fig:scatt}
\end{figure}
The resonance is created by spin-exchange coupling~\cite{Chin2010fri} of the colliding pair to a bound state of the same $M_F = m_1 + m_2$.
Here $m_{1,2}$ are the projections along the magnetic field quantization axis of the total angular momenta of atoms 1 and 2, $\vec{f}_{1,2}$, and $M_F$ is that of the total spin angular momentum, $\vec{F} = \vec{f}_1 + \vec{f}_2$.
We note that $f$ and $F$ are only good quantum numbers at zero magnetic field.

For a pair of atoms in an excited Zeeman channel, there are two processes that can cause two-body collisional loss.
Spin-exchange coupling can lead to inelastic spin relaxation (ISR), in which the colliding pair is coupled into an energetically lower channel of the same $M_F$ and the same partial wave $\ell$.
Since each resonance considered in this work is in the energetically lowest $s-$wave channel of the relevant $M_F$, ISR does not occur.
Spin dipole coupling~\cite{Chin2010fri}, however, can couple a colliding pair to channels
of a different $M_F$ or $\ell$, under the constraints that $M_\p{tot} = M_F + m_\ell$ is conserved, and the change in partial wave is given by $\Delta \ell = \pm 2$.
Here,  $m_\ell$ is the projection of $\vec{\ell}$ along the magnetic field quantization axis.
For the resonances considered here, spin dipole coupling links $s$- and $d$-waves, $d$- and $g$-waves, etc., with odd partial waves excluded by symmetry requirements.
For the $ac$ channel, the two main decay pathways are the $aa$ and $ab$ $d-$wave channels.
Consequently, a basis including all $s-$ and $d-$wave channels with $M_\p{tot} = -2$ is sufficient.

Scattering properties in the vicinity of the resonance at 154.75\,G are shown in \reffig{fig:scatt}, along with the fit from Eqs.~(\ref{eq:K2b})-(\ref{eq:bofb}).
The calculation used a collision energy of $k_\p{B} \times 1\,$pK, while the fit assumes zero temperature.
The fit gives excellent agreement, with only small deviations visible outside the core of the resonance, and at the very center, where effects related to finite collision energies become important.
The background scattering length near the resonance is $63.0\,a_0$, a suitable value for evaporative cooling, while the resonance width of 0.88\,G makes it easily accessible experimentally.
Suppression of collisional losses is provided by the $k_\p{B} \times 14$\,mK height of the $d$-wave barrier being greater than the Zeeman splitting ($k_\p{B} \times 1.8$\,mK for $ab$, $k_\p{B} \times 3.6$\,mK for $aa$) between the entrance and exit channels.
Consequently, decaying pairs must tunnel through the centrifugal barrier.
The resulting resonance length is $4.0 \times 10^6 \, a_0$.
This is comparable to results we have found for much broader, entrance-channel dominated resonances, such as the $ee$ resonance of $^{85}$Rb at $B_0 = 155\,$G, which has $a_\p{res} = 2.5 \times 10^6 a_0$.
We note that three-body effects, not included in our calculations, are also of significance for experiments.

\subsection{Scattering properties: Experiment}
\label{sec:scexpts}

\subsubsection{Experimental conditions}

The basic procedures to prepare the 
Fermi-Fermi mixture near the 155\,G Feshbach resonance are described in Ref.~\cite{Spiegelhalder2010aopNOTE}. Here we briefly summarize the main experimental parameters, and mention some issues of particular relevance for the present experiments.

Our optical dipole trapping scheme employs two stages. In the first stage, we use a high-power laser source (200\,W fiber laser) to load and evaporatively cool the mixture \cite{Spiegelhalder2009cso,Spiegelhalder2010aopNOTE}. As the quality of this high-power beam suffers from thermally induced effects such as spatial shifts and thermal lensing effects, we transfer the mixture into a second trapping beam that uses less laser power and is optimized for beam quality. This beam serves as the trapping beam in the second stage where all the measurements are performed. As the laser source we either use a broadband 5\,W fiber laser (IPG YLD-5-1064-LP, central wavelength 1065\,nm) or a 25\,W single-mode laser (ELS VersaDisk 1030-50, central wavelength 1030\,nm)\footnote{Specific product citations are for the purpose of clarification only, and are not an endorsement by the authors, JQI or NIST.}. In both cases the trapping potential (Gaussian beam waist 41\,$\mu$m) is essentially the same, but we found that the broadband fiber laser can induce inelastic losses \cite{lasersource}. For our measurements on elastic interactions (Sec.~\ref{sec:expelastic}) we used the broadband laser, and we later switched to the single-mode laser for the measurements of inelastic decay (Sec.~\ref{sec:expinelastic}). At a laser power of 70\,mW the trapping frequencies for Li (K) are 13\,Hz (4.5\,Hz) axially \cite{axconfinement} and 365\,Hz (210\,Hz) radially, and the trap depth is 1.6\,$\mu$K (3.6\,$\mu$K).

The mixture contains about $10^{5}$ Li atoms at a temperature $T^{\rm Li}\approx 140$\,nK together with about $2\times 10^{4}$ K atoms at a temperature $T^{\rm K}\approx 160$\,nK; the temperatures are measured by time-of-flight imaging. Note that the two species are not fully thermalized at this point, such that $T^{\rm K}>T^{\rm Li}$. In terms of the corresponding Fermi temperatures $T_F^{\rm Li}=490$\,nK and $T_F^{\rm K}=140$\,nK, the temperatures can be expressed as $T^{\rm Li}/T_F^{\rm Li} \approx 0.3$ and $T^{\rm K}/T_F^{\rm K} \approx 1.1$.

The magnetic field was calibrated by driving RF transitions between the $b$ and the $c$ state of K and using the Breit-Rabi formula \cite{Bcalib}.
In our set of measurements on the elastic scattering properties (Sec.~\ref{sec:expelastic}) the magnetic-field uncertainty was about 20\,mG, with a substantial contribution from magnetic field ripples connected with the 50-Hz power line. In the later experiments on inelastic decay (Sec.~\ref{sec:expinelastic}) we could reduce this uncertainty down to about 5\,mG.

\subsubsection{Elastic Scattering}
\label{sec:expelastic}

Our measurements on elastic scattering are based on the observation of sloshing motion, serving as a simple and sensitive probe for interspecies interactions \cite{Gensemer2001tfc,Maddaloni2000coo,Ferrari2002cpo,Ferlaino2003boi}. Without interaction both components would oscillate independently with their different sloshing frequencies. The interaction induces friction between the two components and thus leads to damping. In the regime of weak interactions with up to a few scattering events per oscillation period, the damping rate can be assumed to be proportional to the elastic scattering cross section. Note that an alternative approach, based on cross-dimensional relaxation, was followed in Ref.\ \cite{Costa2010swi}.

Here we restrict our attention to the slow axial sloshing motion. We excite this motion by an additional infrared beam intersecting our trapping beam \cite{displacebeam}. The magnetic field is quickly ramped to the final setting that is applied in the measurements. After a variable hold time in the trap, we image both clouds to record their damped oscillatory motions. Our data analysis is based on the position of the K cloud, which is completely immersed in the much larger Li cloud. Its motion is analyzed by fitting a simple damped harmonic oscillation
\begin{equation}\label{eq:dampedOsc}
z(t) = Ae^{-\zeta t}\sin{\left(\omega t +  \phi\right)} + z_0
\end{equation}
to the observed axial center-of-mass position. Here $A$ is the oscillation amplitude, $\zeta$ represents the damping rate, 
$\omega$ is the oscillation frequency, and $z_0$ the equilibrium position.

\begin{figure}[tb]
\includegraphics[width=0.95\columnwidth]{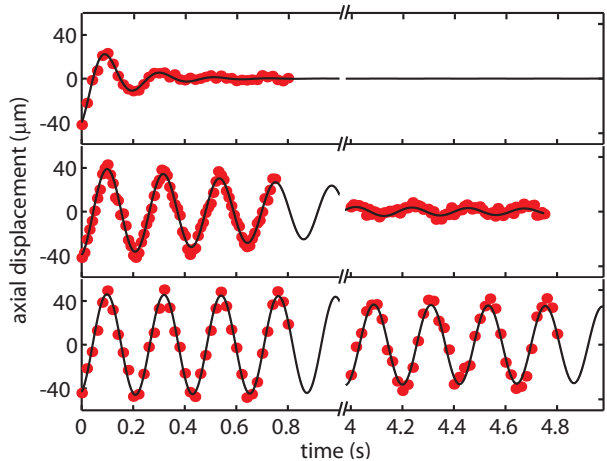}
\caption{\label{fig:DampingSamples} (Color online) Samples of the damped sloshing motion of $^{40}$K for three different settings of the magnetic field (upper panel 154.60\,G, middle panel 154.96\,G, lower panel 155.66\,G). The solid circles represent the experimental data, with uncertainties smaller than the size of the symbols. The solid lines are fits under the assumption of a simple damped harmonic oscillation.}
\end{figure}

\begin{figure}
\includegraphics[width=0.95\columnwidth]{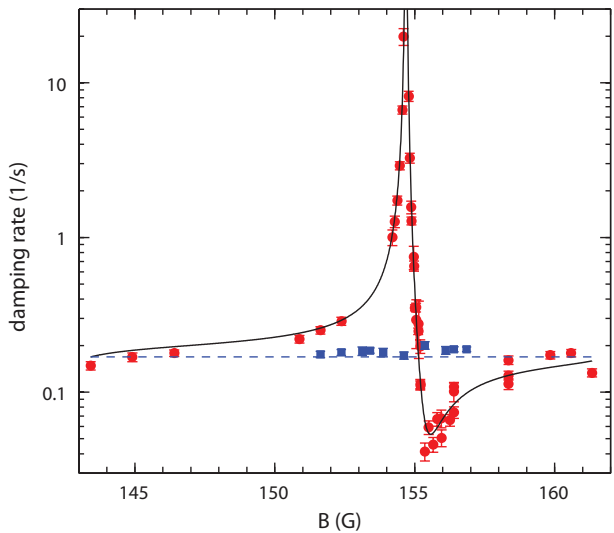}
\caption{\label{fig:Fig1FeshbachScan} (Color online) Elastic scattering near the 155\,G interspecies Feshbach resonance.
The measured rates for the $^{40}$K sloshing motion are shown (filled red circles) together with a fit based on a coupled-channels calculation of the scattering length (solid black line). For comparison we also plot damping rates measured for the non-resonant $ab$ channel (solid blue squares) together with a theoretical line derived from the corresponding non-resonant scattering length (dashed blue line). The error bars shown for the experimental data indicate the fit errors of the damping rate.}
\end{figure}

Near the Feshbach resonance, the observed damping strongly depends on the magnetic field, as demonstrated by the three sample oscillations displayed in Fig.~\ref{fig:DampingSamples}. The measured damping rate as a function of the magnetic field, shown in Fig.~\ref{fig:Fig1FeshbachScan}, reflects the characteristic Fano profile of the elastic scattering cross section. The measured rates vary over three orders of magnitude, prominently showing both the pole of the resonance and its zero crossing.

We analyze the observed magnetic-field dependence of the damping under the basic assumption that the rate $\zeta$ is proportional to the elastic scattering cross section, which itself is proportional to $a^2$. Moreover, we take a background damping \cite{backgrounddamp} into account which is independent of the interspecies interaction and express the total magnetic-field dependent damping rate as
\begin{equation}
\zeta(B) = A a(B)^2 + \zeta_0 \,.
\label{eq:zetaB}
\end{equation}
For the scattering length $a(B)$ we use the result of the coupled-channels calculation as presented in Sec.\ \ref{sec:sctheory}.
The theory has an uncertainty in the resonance position of the order of 100\,mG, limited by the accuracy of the spectroscopically derived potentials.
We therefore allow for a magnetic-field offset by setting
\begin{equation}
a(B) = a_{\mathrm cc}(B + \delta) \,,
\label{eq:shift}
\end{equation}
where $a_{cc}(B)$ refers to the scattering length resulting from the coupled-channels calculation (Sec.~\ref{sec:sctheory}) and $\delta$ is used as a free parameter. Based on Eqs.~(\ref{eq:zetaB}) and (\ref{eq:shift}) we fit the experimental damping data with the three free parameters $A$, $\zeta_0$, and $\delta$.

The fit result, shown by the solid line in Fig.~\ref{fig:Fig1FeshbachScan}, shows excellent agreement with the experimental data. For the background damping of the non-interacting mixture, the fit yields $\zeta_0 = 0.0053(3)$\,s$^{-1}$, which is consistent with independent measurements on K without Li \cite{backgrounddamp}. For the magnetic field offset parameter, the fit yields $\delta = +69(3)$\,mG. Based on the theoretical results of Sec.~\ref{sec:sctheory} and this shift, we obtain a resonance position of 154.69(2)\,G with the 20\,mG calibration uncertainty being the dominant error source.

The experimental data can also be analyzed independently of the coupled-channels calculations by using the standard Feshbach resonance expression (Eq.\ (\ref{eq:aofb}) in the limit $\gamma_\p{B} \rightarrow 0$) for a fit in which the width $\Delta$ is kept as a free parameter. Our corresponding result $\Delta = 920(50)$\,mG is consistent with the prediction $\Delta = 880$\,mG resulting from the coupled-channels calculation.

For comparison, we have also investigated elastic scattering in a Li-K mixture in the $ab$ channel (solid squares in Fig.~\ref{fig:Fig1FeshbachScan}), which near 155\,G is weakly interacting. Our measurements show a damping of the sloshing motion that is consistent with the predicted non-resonant scattering length of 68\,$a_0$ for this channel (solid line).

\subsubsection{Inelastic Scattering}
\label{sec:expinelastic}

To probe inelastic decay, we first prepare a weakly interacting, long-lived Li-K mixture in the $ab$ channel at a variable magnetic field near 155\,G. We then apply a short RF $\pi$-pulse (duration 90\,$\mu$s) to quickly transfer the mixture into the $ac$ channel. This transfer method avoids fast magnetic field ramps and thus any waiting time for the magnetic field to be settled to a constant value before measurements can be taken.

\begin{figure}
\includegraphics[width=0.95\columnwidth]{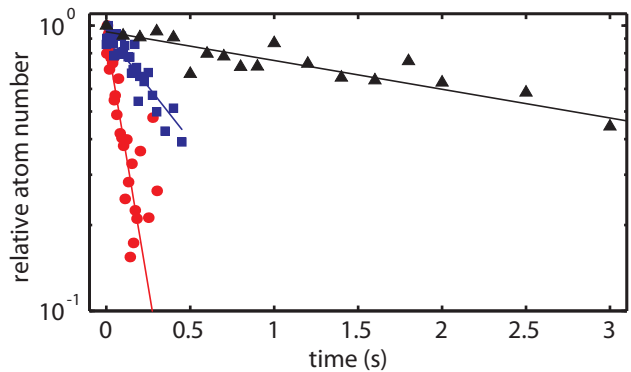}
\caption{\label{fig:FigLifetimeSamples} (Color online) Decay of K immersed in the Li cloud for different magnetic fields, far away from resonance at 154.766\,G (triangles), very close to resonance at 154.712\,G (circles), and for a setting in between at 154.727\,G (squares). The scatter of the data indicates the shot-to-shot variations of the measurements. As the lifetimes are plotted on a logarithmic scale, the good linear fits indicate pure exponential decay.}
\end{figure}

Figure~\ref{fig:FigLifetimeSamples} shows example decay curves. The K loss is essentially exponential, which results from the fact that the K cloud is immersed in a much larger Li sample \cite{Spiegelhalder2009cso}. In this regime, the Li cloud serves as a large bath with essentially constant density. Here the loss curves do not allow us to distinguish between two-body losses where one K atom interacts with one Li atom and such three-body losses, where one K atom interacts with two Li atoms. Three-body losses resulting from two K atoms interacting with one Li atom would not lead to exponential loss.

We analyze the decay under the hypothesis of dominant two-body loss, which is motivated by the decaying character of the Feshbach resonance as discussed in Sec.~\ref{sec:sctheory}. The total K decay rate can be approximated by $\Gamma = K_2 \langle n_{\rm Li} \rangle + \Gamma_{\rm bg}$,
where $\Gamma_{\rm bg}$ is a small background loss rate \cite{bgloss}. The mean Li number density is given by $\langle n_{\rm Li} \rangle$, where the angle brackets denote averages weighted by the K density distribution. For our experimental parameters we obtain $\langle n_{\rm Li} \rangle = 5.9 \times 10^{11}$\,cm$^{-3}$, which is about 75\,\% of the peak density in the center of the Li cloud.

\begin{figure}
\includegraphics[width=0.95\columnwidth]{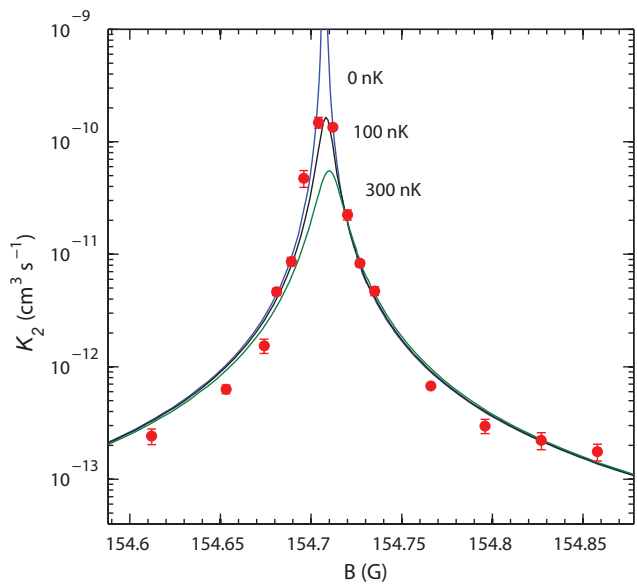}
\caption{(Color online) Inelastic loss near the 155\,G interspecies Feshbach resonance. The measured values for the two-body loss rate coefficient $K_2$ (solid circles) are compared with theory. The three theoretical curves (solid lines) represent three different collision energies $E/k_\p{B}$ (1\,pK, labelled 0 to indicate the zero energy limit, 100\,nK, and 300\,nK), showing the limiting effect of finite collision energy. The error bars represent the statistical errors from fitting the loss curves.}
\label{fig:FigK2Coeff}
\end{figure}

Figure \ref{fig:FigK2Coeff} shows the measured values for the loss rate coefficient $K_2$. The data show the expected resonance behavior (Sec.\ \ref{sec:sctheory}). The values peak at the center of the Feshbach resonance and strongly decrease within a few 10\,mG away from the center. For large scattering lengths, the data follow the scaling $K_2 \propto a^2$ according to Eq.\ (\ref{eq:K2a2}). The observed resonance behavior thus confirms our assumption of the dominant two-body nature of loss. Three-body losses in a two-component Fermi mixture would show a much stronger dependence on $a$ \cite{Dincao2005sls}, not consistent with this observed behavior. However, significant three-body loss contributions may be present very near to the resonance center.

The figure also shows three theoretical curves, calculated for three different values of the collision energy ($k_\p{B} \times 1$\,pK, representing the zero energy limit, $k_\p{B} \times 100$\,nK, and $k_\p{B} \times 300$\,nK) in a range relevant for our experiments.
As a typical value for the collision energy, we can consider an estimate of $200$\,nK \cite{collenergy}.
Very close to the resonance the theory curves illustrate how $K_2$ increases in the zero temperature limit up to a value corresponding to $b=2 a_\p{res}$. In the case of non-zero collision energies it is limited to lower values. For magnetic detunings exceeding about 20\,mG, the effect of the finite collision energies can be neglected in the interpretation of the experimental data, which makes the comparison between theory and experiment straightforward. Here we find excellent quantitative agreement, confirming two-body decay as the dominant loss mechanism. Very close to the center of the resonance the situation is more complicated. If one completely attributes loss to two-body decay, the $100$\,nK curve provides an excellent fit to the experimental data. This, however, is somewhat below our estimate of $200$\,nK for an effective collision energy, which may point to additional three-body losses at the very center of the resonance.

To extract the precise resonance position we proceed in an analogous way as for analyzing the elastic scattering data, allowing for a small magnetic field shift $\delta$ between theory and experiment. We write the actual loss coefficient as $K_2(B) = K_{\mathrm 2,\,cc}(B + \delta)$, where $K_{\mathrm 2,\,cc}$ refers to the coupled-channels result for $K_2$ as discussed in Sec.\ \ref{sec:sctheory}. In the fit, we exclude the three experimental data points that exceed $3 \times 10^{-11}$\,cm$^{3}/s$ to avoid the region where finite collision energies become important. This also makes sure that the loss data are dominated by two-body decay. The shift $\delta$ is the only free parameter, and we obtain a small value of $\delta = +38(1)$\,mG, well in the range of the theoretical uncertainty.
We finally obtain a resonance position of $B_0 = 154.707(5)$\,G, where the main uncertainty results from the magnetic field calibration. Within the experimental uncertainties this value is consistent with the less precise resonance position obtained from elastic scattering measurements.

\subsection{Bound state properties}
\label{sec:boundstate}

\begin{figure}
	\centering
	\includegraphics[width=0.9\columnwidth, clip]{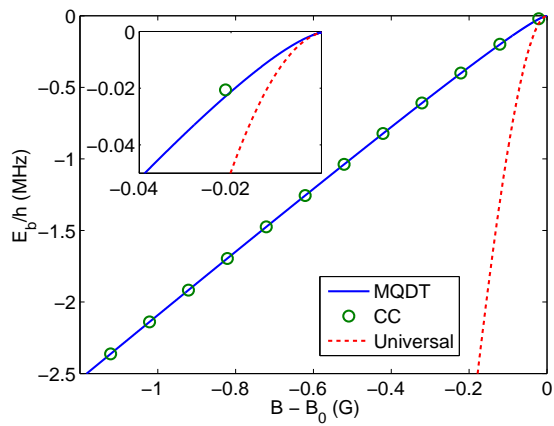}\\
	\caption{(Color online) Molecular binding energy as a function of magnetic field. MQDT (blue line) refers to the three-parameter model of Ref.~\cite{Hanna2009pof}, while CC (green points) indicates a coupled channels calculation. Sufficiently close to the resonance the binding energy converges to the universal result of \refeq{eq:eb_universal} (red dashed line).}
	\label{fig:eb}
\end{figure}
In the context of Feshbach molecules, universality refers to the range of magnetic fields sufficiently close to resonance within which the molecular and scattering properties can be described solely by the atomic masses and the scattering length $a(B)$.
Within this region, the molecule has the form of a halo state, in which a significant part of the wavefunction lies beyond the classically allowed outer turning point of the potential.
This results in a strong enhancement of the lifetime of a decaying bound state~\cite{Thompson2005sdo,Kohler2005sdo}.
The universal binding energy is given by
\begin{equation}
E_\p{B} = \frac{\hbar^2}{2\mu a(B)^2} \,  .
\label{eq:eb_universal}
\end{equation}
Calculations of the relevant bound state energy using the coupled channels method and the simplified three-parameter model of Ref.~\cite{Hanna2009pof} are shown in \reffig{fig:eb}.
We note that the three-parameter model, while useful for bound state and resonance characterisation, does not couple partial waves and so can not be used for calculating decay properties in the present case.
The energy variation is linear for binding energies greater than a few tens of kHz, having a relative magnetic moment of $\delta\mu/h = 2.3$\,MHz/G with respect to the $ac$ threshold.
The universal region, as can be seen from the inset of \reffig{fig:eb}, covers a magnetic field range of order mG.
This makes it hard to access experimentally.
The universal region is wider for broad, entrance channel dominated resonances~\cite{Chin2010fri}.
However, in the present case, the suppression of decay by the centrifugal barrier allows the molecules to have a long lifetime in the nonuniversal regime.

We now consider the lifetime of $^{6}$Li-$^{40}$K molecules close to the $ac$ resonance at 155\,G.
Outside the very narrow universal region, the analytic approach of Ref.~\cite{Kohler2005sdo} does not apply.
We therefore derive the molecular lifetime from a coupled channels scattering calculation including the two open $d$-wave channels into which it decays.
The spin-dipole induced decay discussed in the previous section is mediated by the bound state causing the resonance.
Around the bound state energy, the off-diagonal $T$ matrix element between the two decay channels follows the form
\begin{equation}
|T_{12}|^2 = \frac{\hbar^2\gamma_{1} \gamma_{2}}{(E - E_\p{res})^2 +
\hbar^2 \left( \frac{\gamma_{1} + \gamma_{2}}{2} \right)^2 }\, .
\label{eq:t12}
\end{equation}
Here, $\gamma_{1}/2$ and $\gamma_{2}/2$ are the decay rate of the molecule into the $ab$ and $aa$ channels, respectively.
The total decay rate is given by $(\gamma_{1} + \gamma_{2})/2$.
This calculation includes the entrance channel component and so reproduces the increase in lifetime as the Feshbach molecule takes on a halo form.
This should be distinguished from the decay rate of the bare resonance state which appears in Eqs.~(\ref{eq:aofb}) and (\ref{eq:bofb}).

\begin{figure}
	\centering
	\includegraphics[width=0.9\columnwidth, clip]{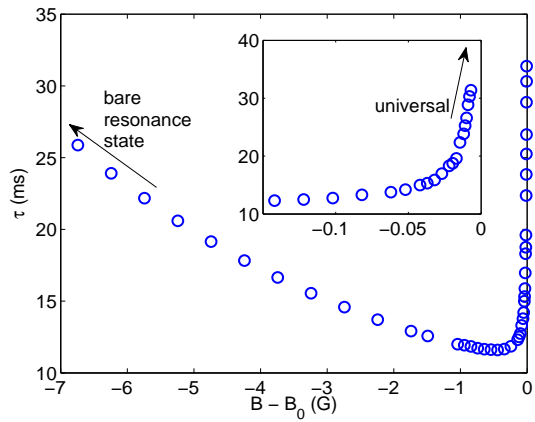}\\
	\caption{(Color online) Calculated molecular lifetime as a function of magnetic field. The molecular lifetime varies at lower field according to its tunnelling rate through the $d$-wave centrifugal barrier into the exit channels. The arrows indicate the sharp increase in lifetime as the universal regime is entered, and the region away from resonance where the molecular lifetime converges to that of the bare resonance state.}
	\label{fig:lifetime}
\end{figure}

Our calculated lifetimes are shown in \reffig{fig:lifetime}.
As discussed above, a sharp increase in lifetime occurs as the universal region near $B_0$ is approached.
Above the maximum lifetime shown in the \reffig{fig:lifetime}, the decay peak described by \refeq{eq:t12} narrows to the point where we can no longer resolve it in our calculations.
The slower increase as $B$ is moved away from $B_0$ occurs because the bound state moves further behind the centrifugal barrier.
Decay from tunnelling through the barrier is then further suppressed.
The lifetime of molecules in the vicinity of the 155\,G resonance was measured by Voigt \textit{et al.}~\cite{Voigt2009uhf}.
They observed a sharp increase in lifetime near the resonance, with which our results qualitatively agree.
Their measured background lifetime of $\sim\!3$\,ms away from resonance is lower than our calculated minimum of 11\,ms.
However, our calculations do not include relevant atom-dimer and dimer-dimer collisions, and so may be considered as an upper bound to experimentally observable lifetimes.
A lifetime of several ms will permit measurements and manipulation of the Feshbach molecules.

\section{Survey of resonances}
\label{sec:survey}

\begin{figure}
	\centering
\includegraphics[width=0.9\columnwidth, clip]{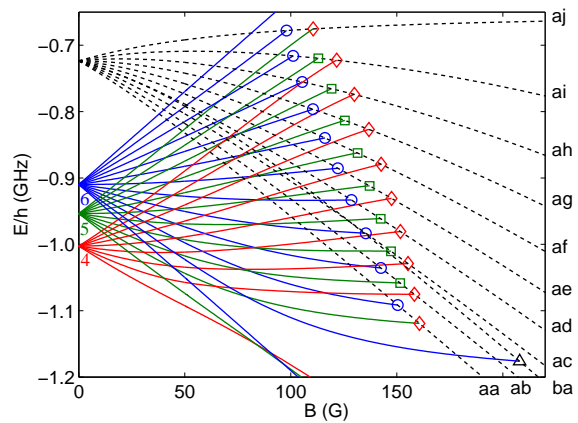}
	\caption{(Color online)
	Energies of the bound states underlying the resonances studied in this paper, as a function of magnetic field.
	Collision thresholds of relevant channels are shown as dotted lines, labelled at the right of the figure.
	The resonances arise from three zero-field bound states of $F = 4, 5$ and $6$ (labelled at the left of the figure), which are projected into channels of $|M_\p{tot}| \leq F$ at non-zero field.
	Deeper bound states are not shown for reasons of clarity.
	Avoided crossings between bound states of the same $M_\p{tot}$ give rise to three identifiable groups of resonances, indicated by the same symbols used in \reffig{fig:survey} and Table~\ref{tab:survey}.
	Within each group, resonance parameters vary smoothly (see Fig.~\ref{fig:survey} and text).
	}
	\label{fig:boundstatemap}
\end{figure}
In this section we discuss resonances occurring in various channels of the $^{6}$Li-$^{40}$K mixture.
We focus on channels with $^{6}$Li in the $a$ state and $^{40}$K in the lower ($f = 9/2$) manifold, for which inelastic spin-exchange collisions do not occur.
At zero magnetic field, there are three bound states of $F = 4,\,5$ and $6$ in the range 200\,MHz to 300\,MHz below these thresholds, as shown in \reffig{fig:boundstatemap}.
At nonzero magnetic field, these states are projected into their Zeeman sublevels, which give rise to Feshbach resonances when degenerate with the collision threshold of a channel of the same $M_\p{tot}$.
Consequently, three proximate resonances are found in channels of $-3 \leq M_\p{tot} \leq 4$.
The bound state underlying each resonance adiabatically correlates with one of the zero field $F$ states.
We note that the bound state energies shown in \reffig{fig:boundstatemap} were produced with the three-parameter model of Ref.~\cite{Hanna2009pof}, which produces slightly different resonance locations to the coupled channels calculations that follow.

\begin{figure}
	\centering
\includegraphics[width=0.9\columnwidth, clip]{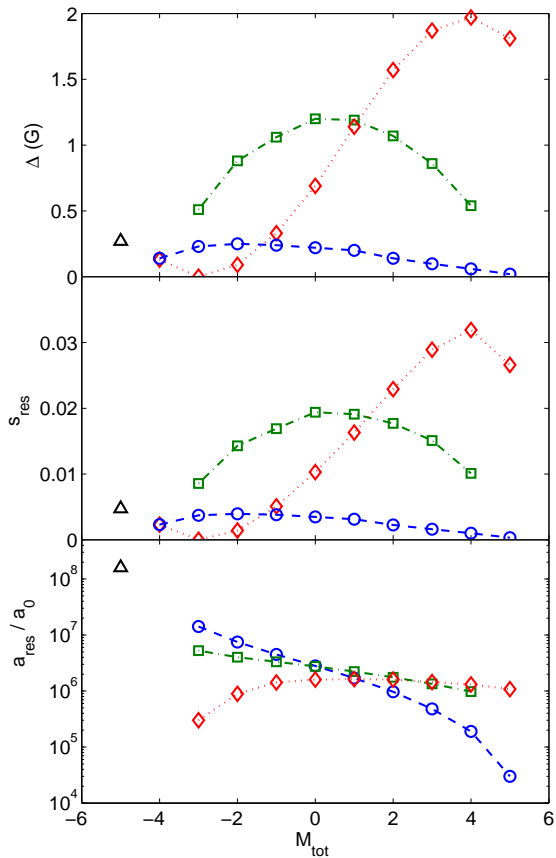}
	\caption{(Color online)
	Theoretical survey of $^{6}$Li--$^{40}$K $s$-wave resonances. Each panel shows a resonance parameter as a function of $M_\p{tot}$: the width $\Delta$ (top), strength $s_\p{res}$ (middle), and length $a_\p{res}$ (bottom).
	The $^{6}$Li atom is in the $a$ state, except for $M_\p{tot} = -5$ for which it is in $b$. The $^{40}$K atom is in the lowest Zeeman state producing the relevant $M_\p{tot}$. The symbols, also used in \reffig{fig:boundstatemap} and Table~\ref{tab:survey}, correspond to the resonance groups discussed in the text. }
	\label{fig:survey}
\end{figure}


\begin{table*}
\begin{center}
\begin{tabular}{| c c c | c c c | c c  c c |c c c|}
\hline\T
&  & & \multicolumn{3}{c|}{Experiment} & \multicolumn{7}{c|}{Coupled channels} \\
\hline\T	
Channel & $M_\p{tot}$ & Group & $B_0$	& $\Delta$  & Ref.	& \:\:\:\:$B_0$\:\:\:\:	& $\Delta$ & $a_\p{bg} / a_0$ & $\delta\mu/h$ & $a_\p{res}$ & $s_\p{res}$  & $\gamma_B$ \\
  & & & (G) & (G) & & (G) & (G) & & (MHz/G) &($10^{6}\,a_0$) & & ($\mu$G) \\
\hline\T
$ba$ 	& -5	& $\bigtriangleup$ 	& 215.6			& 		& \cite{Wille2008eau}		& 215.52	& 0.27			 & 64.3	& 2.4 	 & 160	 & 0.0048 				& 0.11	\\
\hline\T
$aa$ 	& -4	& \bcirc	&157.6			&  		& \cite{Wille2008eau} 		 & 157.50	& 0.14			& 65.0 	 & 2.3 	 &		& 0.0023 				 & 0	 \\
  		&	& \reddia	& 168.170(10)		& 		& \cite{Spiegelhalder2010aopNOTE}	& 168.04	& 0.13			 & 63.4 	 & 2.5	 	&		& 0.0023 				& 0 \\
\hline\T
$ab$ 	& -3	& \bcirc 	& 149.2			&  		& \cite{Wille2008eau}		 & 149.18	& 0.23			& 67.0 	 & 2.1		 & 14	 	& 0.0037 				 & 1.1	\\
  		&	& \gbox	& 159.5			&  		& \cite{Wille2008eau}		& 159.60	& 0.51			& 62.5 	& 2.4		 & 5.3		& 0.0086 				 & 6.1 \\
		&	& \reddia	& 165.9			&  		& \cite{Wille2008eau}		 & 165.928	& $2 \times 10^{-4}$	 & 58		 & 2.5 	& 0.3 	& $ 3.3 \times 10^{-6}$ 	& 0.04 	\\
\hline\T
$ac$ 	& -2	& \bcirc 	& 141.7			& 		& \cite{Wille2008eau} 		 & 141.46	& 0.25			& 67.6 	 & 2.1		 & 7.5	 	& 0.0040 				 & 2.3 \\
  		&	& \gbox	& 154.707(5)		&  	0.92(5)	& this work			& 154.75	& 0.88			& 63.0 	& 2.3	 	 & 4.0	 	& 0.014 				 & 14 	 \\
		&	& \reddia	& 162.7			& 		& \cite{Wille2008eau}		 & 162.89	& 0.09			& 56.4 	 & 2.5		 & 0.89  	& 0.0014 				 & 5.7  \\
\hline\T
$ad$ 	& -1	& \bcirc 	& 				& 		&  						 & 134.08	& 0.24			& 68.7 	 & 2.0		 & 4.5 	& 0.0038 				 & 3.7 \\
  		&	& \gbox	& 				& 		&						& 149.40	& 1.06			& 63.8 	& 2.2		 & 3.3		& 0.017 				 &  20 \\
		&	& \reddia	& 				&  		& 						& 159.20	& 0.33			& 55.8 	& 2.45	 & 1.4		 & 0.0051 				 & 13 \\
\hline\T
$ae$ 	& 0	& \bcirc 	& 	 			&  		&						& 127.01	& 0.22			& 68.5 	& 2.05	 & 2.8		 & 0.0035 				 & 5.4 \\
  		&	& \gbox	& 				& 		& 						& 143.55	& 1.20			& 65.7 	& 2.2		 & 2.8		& 0.020 				 & 29 \\
		&	& \reddia	& 				&  		&						& 154.81	& 0.69			& 55.1 	& 2.4		 & 1.6		& 0.010 				 & 24 \\
\hline\T
$af$ 		& 1	& \bcirc 	& 				& 		& 							 & 120.33	& 0.20			& 66.8 	 & 2.1		 & 1.7		& 0.0031 				 & 7.9 \\
  		&	& \gbox	& 				&	 	& 						& 137.23	& 1.19			& 65.3 	& 2.2		 & 2.2 	& 0.019 				 & 35 \\
		&	& \reddia	& 				& 		&						& 149.59	& 1.14			& 53.6 	& 2.4		 & 1.6		& 0.016 				 &  37 \\
\hline\T
$ag$ 	& 2	& \bcirc 	& 				&  		& 						& 114.18	& 0.14			& 67.4 	 &2.1		 & 0.97	& 0.0023				 & 9.7 \\
   		&	& \gbox	& 				& 		& 						& 130.49	& 1.07			& 66.4 	& 2.2		 & 1.8		& 0.018				 & 40 \\
		&	& \reddia	& 				&  		& 						& 143.39	& 1.57			& 54.4 	& 2.4		 & 1.6		& 0.023				 & 53 \\
\hline\T
$ah$ 	& 3	& \bcirc 	& 				& 		&  						& 108.67	& 0.098			& 66.6 	& 2.2		 & 0.48	& 0.0016 				 & 14 \\
  		&	& \gbox	& 				& 		& 	 					& 123.45	& 0.86			& 68.4 	& 2.3		 & 1.3		& 0.015 				 & 44 \\
		&	& \reddia	& 				&  		& 						& 135.90	& 1.87			& 55.9 	& 2.45	 & 1.5 	& 0.029 				& 72 \\
\hline\T
$ai$ 		& 4	& \bcirc 	&				& 		& 							 & 104.08	& 0.06			& 65.9 	 & 2.25	& 0.19	& 0.0010 				 & 21 \\
 		&	& \gbox	& 				& 		& 						& 116.38	& 0.54			& 68.6 	& 2.4		 & 0.98	& 0.010 				 & 38 \\
 		&	& \reddia	& 				& 		& 						& 126.62	& 1.97			& 54.7 	& 2.6		 & 1.3 	& 0.032 				 & 83 \\
\hline\T
$aj$ 		& 5	& \bcirc 	& 				& 		&  							 & 100.90	& 0.02			& 64.3 	 & 2.3		& 0.03	& $3.2 \times 10^{-4}$ 	& 43 \\
 		&	& \reddia	& 114.47(5)		& 1.5(5)	& \cite{Tiecke2010bfr}		 & 114.78	& 1.81			& 57.3 	 & 2.3		 & 1.08	& 0.027 				 & 96 \\
\hline
\end{tabular}
\end{center}
\caption{Survey of $s$-wave resonances in $^{6}$Li-$^{40}$K. The first two columns give the channel $\alpha\beta$ and total angular momentum projection $M_\p{tot}$, with $\alpha$ and $\beta$ representing the Zeeman state of Li and K, respectively. The third column gives the symbol for the corresponding group of resonances that is used in Figs.~\ref{fig:boundstatemap} and \ref{fig:survey}. The next three columns give experimental values of the resonance location $B_0$ and width $\Delta$, where available, with references. The remaining columns give the results of the coupled channels calculations performed for the current work - $B_0$ and $\Delta$, as well as the background scattering length $a_\p{bg}$, relative magnetic moment $\delta\mu$, resonance length $a_\p{res}$, resonance strength $s_\p{res}$, and decay rate in magnetic field units, $\gamma_B$. Note that $a_\p{res}$ is not defined for the stable $aa$ channel. Note also that the experimental values for $B_0$ from Ref.\ \cite{Wille2008eau} are subject to typical uncertainties of about 0.5\,G.
}
\label{tab:survey}
\end{table*}

We have performed a coupled channels analysis of each resonance, analogous to that performed with the asymptotic bound state model in Ref.~\cite{Tiecke2010bfr}.
With our more rigorous approach \cite{Tiemann}, we obtain good agreement with all experimental data, including the new set of measurements on the 155\,G resonance presented in Sec.\ \ref{sec:scexpts}. The simplified approach of Ref.~\cite{Tiecke2010bfr} seems to underestimate the widths of the resonances by almost a factor of two.
With the coupled channels approach, we can also study the decay properties of the resonances.
The resonance parameters are shown in \reffig{fig:survey}, and tabulated in Table \ref{tab:survey}.
We group the resonances of each channel that are lowest (\bcirc), middle (\gbox) and highest (\reddia) in $B_0$, using the indicated symbols to distinguish the resonance groups in Figs.~\ref{fig:boundstatemap} and \ref{fig:survey}, and Table~\ref{tab:survey}.
Within each of these groups, resonance properties vary smoothly as a function of $M_\p{tot}$.
The resonances with $M_\p{tot} = -4$ and 5 have properties consistent with the lowest and highest group, while the $ba$ resonance with $M_\p{tot} = -5$ has substantially different properties.
This is due to $F$ being a good quantum number only at zero magnetic field, and several bound states having avoided crossings in the relevant range of magnetic field.
%
There are several resonances with $\Delta \gtrsim 1\,$G, offering good opportunities for control of collisional properties. However, several other factors are also useful for deciding the suitability of a resonance for a given application.

One parameter used for quantifying the extent to which a resonance is entrance-channel dominated is the resonance strength parameter~\cite{Chin2010fri}, defined by
\begin{equation}
s_\p{res}  = \frac{a_\p{bg}}{\bar{a}} \frac{\delta \mu \Delta}{\bar{E}} \, .
\end{equation}
Here, $\bar{a} =  [4\pi/\Gamma(1/4)^2] R_\p{vdw} \approx 0.956 R_\p{vdW}$ is the mean scattering length~\cite{Gribakin1993cot}, and $\bar{E} = \hbar^2/(2\mu\bar{a}^2) \approx 1.094 E_\p{vdW}$ is the associated energy.
If $s_\p{res} \geq 1$, the bound state and near-threshold scattering states are concentrated in the entrance channel over a magnetic field range comparable to $\Delta$.
In the present case, all resonances are closed-channel dominated, as shown in the middle panel of \reffig{fig:survey}.
The background scattering lengths of the resonances are all in the range $55\, a_0$ to $70\, a_0$, and the relative magnetic moments are in the range 2\,MHz$/$G to 2.6\,MHz$/$G.
Consequently, the resonance strength follows trends similar to the resonance widths.
For the $ac$ 155\,G resonance we have $s_\p{res} = 0.014$.
This is reflected in the universal region being only a few mG wide, as discussed above.

Calculated resonance lengths are shown in the lower panel of \reffig{fig:survey}.
The range in $a_\p{res}$ is approximately 3 orders of magnitude, with better stability in channels of lower $M_\p{tot}$.
This occurs because the energy gaps between higher channels are larger, reducing the height of the centrifugal barrier through which the decaying atoms tunnel.
However, all the resonance lengths are sufficiently high that we expect each resonance with $\Delta \gtrsim 1$\,G to be very useful.

In view of interaction control in a strongly interacting gas, we now discuss three selected resonances that have received particular attention in experiments: the 168\,G resonance in the $aa$ channel \cite{Spiegelhalder2010aopNOTE}, the 155\,G resonance in the $ac$ channel \cite{Voigt2009uhf} (see Sec.\ \ref{sec:155G}), and the 114\,G resonance in the $aj$ channel. In practice, the possible degree of control is limited by uncertainties (drifts and fluctuations) of the magnetic field. A corresponding figure of merit is the maximum controllable scattering length $a_\p{ctrl} = a_\p{bg} \Delta / \delta_B$, where $\delta_B$ stands for the magnetic field uncertainty. Assuming a realistic value of $\delta_B = 5$\,mG one obtains $a_\p{ctrl} \approx 1600\,a_0$, $11'000\,a_0$, and $21'000\,a_0$ for the three resonances considered (168\,G, 155\,G, and 114\,G, respectively). On one hand, this can be compared with the typical requirement of $|a| \gtrsim 5000\,a_0$ for attaining strongly interacting conditions. On the other hand, it can be compared with the condition for universal behavior $|a| \gg \bar{a}/s_\p{res}$ \cite{Rstar}, which requires $a_\p{ctrl} \gg 17'000\,a_0$, $2800\,a_0$, and $1450\,a_0$. This shows that the resonance at 168\,G is too narrow for controlling a strongly interacting Fermi-Fermi mixture, but the other resonances at 155\,G and 114\,G are broad enough. Although the 114\,G resonance is wider than the 155\,G resonance by a factor of 2.1, inelastic loss is 3.7 times faster. The higher collisional stability is an important advantage of the 155\,G resonance.

\section{Conclusions}
\label{sec:conclusion}

We have characterized elastic and inelastic scattering near Feshbach resonances in the $^6$Li-$^{40}$K mixtures. The presence of open decay channels for all broader resonances has two important consequences. Atomic two-body collisions acquire a resonantly enhanced inelastic component, which unavoidably limits the stability of an atomic Fermi-Fermi mixture with resonantly tuned interactions. When Feshbach molecules are created via these decaying resonances, they will undergo spontaneous dissociation.

The intrinsic decay has important consequences for present experiments towards strongly interacting Fermi-Fermi mixtures. Under typical experimental conditions, the lifetime of a Fermi-Fermi mixture with resonantly tuned interactions ($a \rightarrow \pm \infty$) will be limited to $\sim$10\,ms. This in general means a limitation of possible experiments to short time scales, such as the observation of the expansion of the mixture after trap release \cite{Ohara2002ooa,Bourdel2003mot} or measurements of fast collective oscillation modes \cite{Kinast2004efs,Bartenstein2004ceo,Altmeyer2007pmo}. Experiments that require long time scales, such as precise studies of equilibrium states \cite{Zwierlein2006doo,Nascimbene2010ett}, may be problematic in this decaying mixture.

The short lifetime of the Feshbach molecules, also being of the order of 10\,ms, excludes the production of a long-lived molecular Bose-Einstein condensate (mBEC) such as formed in $^6$Li \cite{Jochim2003bec,Zwierlein2003oob}. Transient ways to form mBECs, as demonstrated for $^{40}$K \cite{Greiner2003eoa}, will still be possible. The detection of fermionic condensates by rapid conversion of many-body pairs into molecules \cite{Regal2004oor} also seems to be a realistic possibility. Moreover, the predicted increase of the molecular lifetime for larger binding energies can be of general interest for the coherent manipulation of Feshbach molecules \cite{Ferlaino2009ufm} and in particular for optimizing the starting conditions for a transfer to the ro-vibrational ground state \cite{Ni2008ahp,Lang2008utm,Danzl2010auh}.

Finally, the question of which Feshbach resonance provides optimum conditions for interaction tuning in the $^6$Li-$^{40}$K has no straightforward answer. All the broad resonances occurring in the channels $ac$\,-\,$aj$ (widths 0.88\,G - 1.97\,G) seem to be well suited for controlled interaction tuning. Because of the tradeoff between width and stability, the best choice will depend on the particular application.

\begin{acknowledgement}
We acknowledge support by the Austrian Science Fund (FWF) and the European Science Foundation (ESF) within the EuroQUAM/FerMix project and support by the FWF through the SFB FoQuS. T.M.H and P.S.J acknowledge support from an AFOSR MURI on Ultracold Molecules.
\end{acknowledgement}


\begin{thebibliography}{69}
\bibitem{Tiemann}
Resonance positions and widths have also been calculated by E. Tiemann using a
  coupled channels approach similar to our theoretical work; private
  communication.
\bibitem{Bcalib}
In the vicinity of the resonance the transition frequency between the $b$ and
  the $c$ state of $^{40}$K can be expressed as $\nu_\p{RF} = 38.5756$\,MHz +
  195.5\,kHz/G $\times$ ($B-154.707\,$G).
\bibitem{displacebeam}
The displacement beam is derived from the same laser source as the trapping
  beam. It has a power of 25\,mW and a waist of $60\,\mu$m, and it intersects
  the trapping beam at an angle of 17.5$^{\circ}$ about $120\,\mu$m away from
  the focus of the trapping beam. By adiabatically turning on the displacement
  beam we shift the centers of the two trapped species in the axial direction.
  Then, by suddenly extinguishing the displacement beam, the clouds are
  released into the unperturbed trap potential and they start their
  oscillations.
\bibitem{backgrounddamp}
Weak damping of K sloshing with a rate of $\sim$0.04\,s$^{-1}$ is also observed
  when the Li component is absent. We attribute this residual damping to
  imperfections of the trapping potential such as corrugations and
  anharmonicities.
\bibitem{collenergy}
The main contribution to the mean collision energy in our trapped sample stems
  from the kinetic energy of the degenerate Li component. In the trap center,
  where the K cloud overlaps with the Li, the mean kinetic energy of the Li
  atoms is given by $(3/10) T^{\rm Li}_{F} \approx 200$\,nK.
\end{thebibliography}

\end{document}